\documentclass[amsmath,aps,pra,twocolumn]{revtex4-1}
\usepackage{graphicx}
\usepackage{times}
\usepackage{slashed}
\usepackage{bm}

\usepackage{color}

\begin{document}
\title{Thermal corrections to the bound-electron $g$-factor}
\author{T. Zalialiutdinov$^{1,\,2}$, D. Glazov$^1$ and D. Solovyev$^1$}

\affiliation{ 
$^1$ Department of Physics, St. Petersburg State University, Petrodvorets, Oulianovskaya 1, 198504, St. Petersburg, Russia\\
$^2$ Special Astrophysical Observatory, St. Petersburg Branch, Russian Academy of Sciences, 196140, St. Petersburg, Russia
}

\begin{abstract}
The influence of the blackbody radiation field on the $g$-factor of light hydrogenlike ions is considered within the framework of quantum electrodynamics at finite temperature for bound states. One-loop thermal corrections are examined for a wide range of temperatures. The numerical results for $1s$, $2s$, $2p_{1/2}$, and $2p_{3/2}$ states are presented. It is shown that for excited states finite temperature corrections to the bound-electron $g$-factor are close to the level of current experimental uncertainty even at room temperatures and can be discerned within the measurements anticipated in the near future.
\end{abstract}

\maketitle
\section{Introduction}

In the presence of external magnetic field, the atomic energy levels are split according to the projection of the total angular momentum relative to the direction of the field --- the so-called Zeeman effect. In the case of weak field, the Zeeman shift is linear in the field strength with the proportionality coefficient termed as $g$-factor. Experiments on the determination of the bound-electron $g$-factor have reached extremely high precision in recent years \cite{werth:18:aamop}. Up to now, the most accurate results were obtained for hydrogen-like ions: $ ^{12}\mathrm{C}^{5+} $ \cite{haefner:00:prl,nature2014}, $ ^{16}\mathrm{O}^{7+} $ \cite{oxygen}, and $^{28}\mathrm{Si}^{13+} $ \cite{sturm:13:pra,nature2014} with the precision on the level of $10^{-12}$. The follow-up measurements in lithium-like \cite{wagner:13:prl,koehler:16:nc,glazov:19:prl} and boron-like \cite{arapoglou:19:prl,egl:19:prl,micke:20:n} ions are evolving towards a comparable level of precision. Analyzing experimental data and corresponding results from elaborate theoretical calculations, one can determine fundamental constants and nuclear parameters \cite{Glazov2014,shabaev:2015:031205,harman:18:jpcs}. In particular, the currently accepted value of the electron mass was obtained from the analysis of the $g$-factor measurements in $ ^{12}\mathrm{C}^{5+} $ and $^{28}\mathrm{Si}^{13+} $ ions and the corresponding theory \cite{nature2014, Mohr-2016, Zatorski}.

To match experimental precision, a rigorous account of relativistic, QED, and nuclear (including recoil, finite size, and polarization) effects is needed \cite{blundell:1997:1857, Persson, lindgrengfactor, recoilgfactor, polgfactor, glazov:2002:408, oneloopgfactor, twoloopgfactor, jentschura:2009:044501, yerokhin:2013:042502, yerokhin:2017:060501, czarnecki:2018:043203, czarnecki:2020:050801}. Even the impact of the gravitational field of the Earth \cite{jentgravitation} or instability of external fields affecting atomic characteristics \cite{Schabinger2012} are investigated. In addition, in view of the achieved experimental and theoretical accuracy, it is necessary to describe in detail the influence of the blackbody radiation field (BBR), which has a Planck equilibrium distribution. It is well known that the BBR field leads to a quadratic ac-Stark shift of energy levels and reduces the lifetimes by inducing electron transitions between atomic states. These effects are extremely important in the spectroscopy of Rydberg atoms, the construction of atomic clocks, and the determination of frequency standards \cite{GC, Hall, AtCl-Sr, AtCl-Cs, Nicholson}. The study of the effect of equilibrium radiation on the characteristics of atomic systems is usually limited to the quantum-mechanical approach, in which the root-mean-squared field induced by BBR is considered as a perturbation. In our recent works, bound-state quantum electrodynamics theory at finite temperature (BS-TQED) has been developed to calculate thermal effects in atomic systems \cite{SLP-QED,S-2020,SZA-2020,SZA_2021}. Within the framework of this theory and Line Profile Approach (LPA) \cite{Andr}, various corrections to the transition probabilities and ionization/recombination cross-sections were also calculated \cite{SLP-QED, jphysb2017, ZSL-1ph, ZAS-2l, SZATL}. In this work, using previously developed methods for calculating finite temperature effects, we study thermal one-loop radiative corrections to the $g$-factor of a bound electron in the presence of BBR. 

The paper is organized as follows. In section \ref{s1} within the framework of rigorous quantum electrodynamics at finite temperature and the adiabatic S-matrix formalism, we derive analytical equations for thermal self-energy radiative corrections to atomic electron $g$-factor. Then the results of the numerical evaluation are discussed in section \ref{s2}. Algebraic transformations used in the calculation are given in Appendix A. The relativistic units $ \hbar=c=m_{e}=1 $ ($ \hbar $ is the Planck constant, $ c $ is the speed of light, $ m_{e} $ is the electron mass, it is written explicitly in some places for clarity) are used throughout the paper. The product of the Boltzmann constant $ k_{B} $ and the temperature $ T $ is written in relativistic units. 

\section{Adiabatic S-matrix formalism for evaluation of thermal radiative corrections}
\label{s1}

We consider a bound-electron state $ |a\rangle \equiv|n_{a}l_{a}j_{a}m_{j_a}\rangle$ characterized by the principal quantum number $ n_{a} $, the orbital quantum number $ l_{a} $, the total angular momentum $ j_{a} $, and its projection $ m_{j_{a}} $. It represents the solution of the Dirac equation for the Coulomb nuclear field with the nuclear charge number $Z$. The energy shift of the atomic level $E_{a}$ in the homogeneous magnetic field $ \textbf{B} $ can be expressed in terms of the four-vector potential $ A_{\mu}^{\mathrm{pert}}=(0,\textbf{A}) $:
\begin{eqnarray}
\label{00}
\Delta E_{a}=-\langle a|e\gamma_{\mu}A_{\mu}^{\mathrm{pert}}|a\rangle = \langle a|e\bm{\alpha}\cdot\textbf{A}|a\rangle
\,,
\end{eqnarray}
where 
\begin{eqnarray}
\label{e3}
\textbf{A}=-[\textbf{r}\times\textbf{B}]/2
\end{eqnarray}
and $\textbf{r}$ is the radius vector for the electron in the atom. 
Then, for the field $ \textbf{B} $ directed along the $z$ axis, we find
\begin{eqnarray}
\label{g1}
\Delta E_{a}=\mu_{B}g^{\mathrm{D}}_{a}B m_{j_a}
\,,
\end{eqnarray}
where $ g^{\mathrm{D}}_{a} $ is termed as Dirac $g$-factor.
For the Coulomb potential of the point-like nucleus, $ g^{\mathrm{D}}_{a} $ can be obtained analytically,
\begin{eqnarray}
\label{corr1}
g^{\mathrm{D}}_{a}=\frac{\kappa_a}{2j_{a}(j_{a}+1)}\left(2\kappa_a\, E_{a}-1\right)
\,,
\end{eqnarray}
where $\kappa_a=l_a(l_a+1)-j_a(j_a+1)-1/4$ is the relativistic angular quantum number, $ E_{a}\equiv E_{n_{a}l_{a}j_{a}} $ is the Dirac energy. In turn, equation (\ref{corr1}) can be presented as expansion in powers of $ \alpha Z $ \cite{Glazov2014} with the zeroth-order term given by the Lande formula (\ref{landeNR}):
\begin{align}
g^{\mathrm{D}}_{1s_{1/2}}=2-\frac{2}{3} (\alpha Z)^2 -\frac{1}{6} (\alpha Z)^4-\dots
\,,\\
g^{\mathrm{D}}_{2s_{1/2}}=2-\frac{1}{6} (\alpha Z)^2 -\frac{5}{96} (\alpha Z)^4-\dots
\,,\\
g^{\mathrm{D}}_{2p_{1/2}}=\frac{2}{3}-\frac{1}{6} (\alpha Z)^2 -\frac{5}{96} (\alpha Z)^4-\dots
\,,\\
g^{\mathrm{D}}_{2p_{3/2}}=\frac{4}{3}-\frac{2}{15} (\alpha Z)^2 -\frac{1}{120} (\alpha Z)^4-\dots
\,.
\end{align}

Introducing the operator of magnetic moment \cite{Persson}
\begin{eqnarray}
\label{fullmu}
 \bm{\mu}=e[\textbf{r}\times\bm{\alpha}]/2
\,,  
\end{eqnarray}
where $\bm{\alpha}$ is the vector of Dirac matrices,
the energy shift $\Delta E_{a}$ (\ref{00}) can be written in the form:
\begin{eqnarray}
\label{0}
\Delta E_{a}=-\langle a|\bm{\mu}\cdot \textbf{B}|a\rangle
\,.
\end{eqnarray}
In the nonrelativistic limit ($ \alpha Z \ll 1 $) \cite{Landau}, the magnetic moment is reduced to
\begin{eqnarray}
\label{munr}
 \bm{\mu}_{\mathrm{nr}}=-\mu_{B}(\textbf{l}+2\textbf{s})=-\mu_{B}(\textbf{j}+\textbf{s})
\,,
\end{eqnarray}
where $ \mu_{B}={|e|}/{(2m_{e})} $ is the Bohr magneton, $ \textbf{l} $ and $ \textbf{s} $ are the operators of orbital momentum and electron spin, $ \textbf{j}=\textbf{l}+\textbf{s} $ is the total angular momentum operator.
The nonrelativistic limit of the Dirac $g$-factor, termed as Lande $g$-factor, is found as
\begin{eqnarray}
\label{landeNR}
g^{\mathrm{L}}_{a} = m_{j_a}^{-1}\langle a | j_z+s_z | a \rangle
\end{eqnarray}
where, $j_z$ and $s_z$ are the $z$-projections of $\textbf{j}$ and $\textbf{s}$, respectively. 

\begin{figure}[hbtp]
\centering
\includegraphics[scale=2]{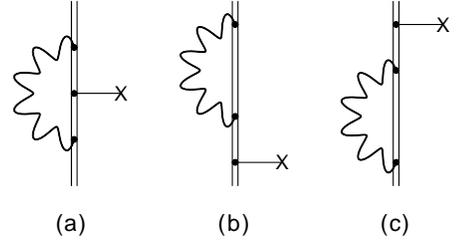}
\caption{Feynman diagrams describing the TQED contributions of the order of $ \alpha $ to the bound-electron $g$-factor. 
The tiny line with the cross indicates interaction with an external magnetic field. The double solid line denotes the bound electron in the Furry picture. The bold wavy line represents the thermal photon propagator.}
\label{fig3}
\end{figure}

From Eq. (\ref{g1}) it follows that an arbitrary correction $ \Delta g_{a} $ to the $g$-factor can be obtained from the corresponding energy shift,
\begin{eqnarray}
\label{corr}
\Delta g_{a}=\frac{\Delta E_{a}}{m_{j_a}\mu_{B}B}
\,.
\end{eqnarray}
Below, we consider radiative corrections for thermal self-energy to the bound-electron $g$-factor in H-like ions via Eq. (\ref{corr}). Within the framework of QED theory, the corresponding energy shift $ \Delta E_{a} $ of the unperturbed values $ E_{a} $ is represented by the three Feynman diagrams in Fig. \ref{fig3}, where, in contrast to the `zero-vacuum' case (see, for example, \cite{Persson}), the ordinary photon propagator is replaced by the thermal one.

Following the Gell-Mann-Low theory (adiabatic $S$-matrix formalism, see \cite{Low, Gell}), the energy correction is given by
\begin{eqnarray}
\label{1}
\Delta E_{a} = \lim\limits_{\eta \rightarrow 0+}\frac{\mathrm{i}\eta }{2}\frac{\frac{\partial}{\partial e }\langle \Phi_a^{0} |\hat{S}_{\eta} |\Phi_a^{0}\rangle}{\langle \Phi_a^{0} |\hat{S}_{\eta} |\Phi_a^{0}\rangle}
\,.
\end{eqnarray}
The evolution operator, $ \hat{S}_{\eta} $, is 
\begin{eqnarray}
\label{2}
\hat{S}_{\eta}=\mathrm{T}\left[\mathrm{exp}\left(-\mathrm{i}e\int d^4x e^{-\eta |t|}\hat{H}_{\mathrm{i}}(x)  \right) \right]
\\\nonumber
=
1+\sum\limits_{k=1}^{\infty}\frac{(-\mathrm{i}e)^k}{k!}\int d^4x_{k}\dots \int d^4x_{1}
\\\nonumber
\times
e^{-\eta |t_{k}|}\dots e^{-\eta |t_{1}|}\mathrm{T}\left[\hat{H}_{\mathrm{i}}(x_{k})\dots \hat{H}_{\mathrm{i}}(x_{1})\right]
\,,
\end{eqnarray}
and $ \mathrm{T}[\dots] $ denotes the time-ordered product of interaction density $ \hat{H}_{\mathrm{i}} $, which is
\begin{eqnarray}
\label{3}
\hat{H}_{\mathrm{i}}(x)=\hat{j}^{\mu}(x)(\hat{A}_{\mu}(x)+\hat{A}_{\mu}^{\mathrm{pert}}(x))
\,.
\end{eqnarray}
Here $ \hat{A}_{\mu}(x) $ and $ \hat{A}_{\mu}^{\mathrm{pert}}(x)) $ are the operators of photon field and external perturbation, respectively.
The operator of electron current in Eq. (\ref{3}) is defined as follows
\begin{eqnarray}
\label{4}
\hat{j}^{\mu}(x)=-\frac{1}{2}[\hat{\overline{\psi}}(x)\gamma^{\mu},\hat{\psi}(x)]
\,,
\end{eqnarray}
where $ \hat{\overline{\psi}}(x)=\hat{\psi}^{\dagger}(x)\gamma_{0} $, $ \hat{\psi}(x) $ is the operator of fermion field and $ \gamma^{\mu} $ are the Dirac gamma matrices.
Since we are interested in the one-loop thermal self-energy contributions, we should take into account the third-order corrections.
Expansion of Eq. (\ref{1}) in powers of $ e $ \cite{LabKlim,LSP-sep} up to the third order yields
\begin{gather}
\label{5}
\Delta E_{a} = \lim\limits_{\eta \rightarrow 0+}\frac{1}{2}\mathrm{i}\eta
\left\lbrace \langle \Phi_a^{0}| \hat{S}_{\eta}^{(1)}|\Phi_a^{0}\rangle
\right.
\\\nonumber
\left.
+
\left[2\langle \Phi_a^{0}| \hat{S}_{\eta}^{(2)}|\Phi_a^{0}\rangle -\langle\Phi_a^{0}| \hat{S}_{\eta}^{(1)}|\Phi_a^{0}\rangle^2\right]
\right.
\\\nonumber
+
\left.
\left[3\langle \Phi_a^{0}| \hat{S}_{\eta}^{(3)}|\Phi_a^{0}\rangle -3
\langle\Phi_a^{0}| \hat{S}_{\eta}^{(2)}|\Phi_a^{0}\rangle
\langle\Phi_a^{0}| \hat{S}_{\eta}^{(1)}|\Phi_a^{0}\rangle 
\right.
\right.
\\\nonumber
+
\left.
\left.
\langle\Phi_a^{0}| \hat{S}_{\eta}^{(1)}|\Phi_a^{0}\rangle^3
\right]
+
\dots
\right\rbrace.
\end{gather}

Here we are dealing with one-electron atomic systems and, therefore, the graphs with more than one fermionic line in the initial and final states are excluded. 
Since we consider the linear effect in magnetic field, we keep only the terms of the first order in the perturbing potential $ \hat{A}_{\mu}^{\mathrm{pert}} $. 




Matrix elements of different orders in $e$ up to $\hat{S}_{\eta}^{(3)}$ are given by the following expressions. Contribution of the first order:
\begin{eqnarray}
\label{6a}
\langle \Phi_a^{0}| \hat{S}_{\eta}^{(1)}|\Phi_a^{0}\rangle
 = (-\mathrm{i}e)\int d^4x\overline{\psi}_{a}(x)
\\\nonumber
\times
e^{-\eta|t|}
\gamma^{\mu}
A_{\mu}^{\mathrm{pert}}(x)
\psi_a(x)
\,.
\end{eqnarray}
The second-order contribution (thermal self-energy):
\begin{eqnarray}
\label{6b}
\langle \Phi_a^{0}| \hat{S}_{\eta}^{(2)}|\Phi_a^{0}\rangle
 = (-\mathrm{i}e)^2\int d^4x_{1}d^4x_{2}\overline{\psi}_{a}(x_{1})\times\qquad
\\\nonumber
e^{-\eta|t_{1}|}
\gamma^{\mu_1}D^{\beta}_{\mu_{1}\mu_{2}}(x_{1},x_{2})\gamma^{\mu_{2}}S(x_{1},x_{2})
e^{-\eta|t_{2}|}
\psi_a(x_{2})
\,.
\end{eqnarray}
The third-order matrix elements, corresponding to the diagrams (a), (b), and (c) in Fig.~\ref{fig3}, are given by
\begin{eqnarray}
\label{6c}
\langle \Phi_a^{0}| \hat{S}_{\eta}^{(3)}|\Phi_a^{0}\rangle_{\mathrm{(a)}}
 = (-\mathrm{i}e)^3\int d^4x_{1}d^4x_{2}d^4x_{3}
 \qquad
\\\nonumber
\times
\overline{\psi}_{a}(x_{1})
e^{-\eta|t_{1}|}
\gamma^{\mu_1}D^{\beta}_{\mu_{1}\mu_{3}}(x_{1},x_{3})e^{-\eta|t_{3}|}\gamma^{\mu_{3}}
\\\nonumber
\times
S(x_{1},x_{2})
e^{-\eta|t_{2}|}
\gamma^{\mu_{2}}A_{\mu_{2}}^{\mathrm{pert}}(x_2)S(x_{2},x_{3})\psi_a(x_{3})
\end{eqnarray}
and
\begin{eqnarray}
\label{6d}
\langle \Phi_a^{0}| \hat{S}_{\eta}^{(3)}|\Phi_a^{0}\rangle_{\mathrm{(b)}} 
= 
\langle \Phi_a^{0}| \hat{S}_{\eta}^{(3)}|\Phi_a^{0}\rangle_{\mathrm{(c)}}
=\qquad
\\\nonumber
(-\mathrm{i}e)^3\int d^4x_{1}d^4x_{2}d^4x_{3}
\overline{\psi}_{a}(x_{1})
e^{-\eta|t_{1}|}
\gamma^{\mu_1}D^{\beta}_{\mu_{1}\mu_{3}}(x_{1},x_{2})
\\\nonumber
\times
e^{-\eta|t_{3}|}
\gamma^{\mu_{3}}S(x_{1},x_{2})
e^{-\eta|t_{2}|}
\gamma^{\mu_{3}}A_{\mu_{3}}^{\mathrm{pert}}(x_3)S(x_{2},x_{3})\psi_a(x_{3}).
\end{eqnarray}
Here
$ \psi_{a}(x)=\psi_{a}(\textbf{r}) e^{\mathrm{i}E_{a}t} $, $ \psi_{a}(\textbf{r}) $ is a solution of the Dirac equation in the state $|a\rangle$ with the energy $ E_{a} $, and $ \overline{\psi}_{a}(x)=\psi_{a}^{\dagger}(x)\gamma_{0} $. The electron propagator $S(x_1,x_2)$ is 
\begin{eqnarray}
\label{7}
S(x_{1}x_{2})=\frac{\mathrm{i}}{2\pi}\int\limits_{-\infty}^{+\infty}d\Omega\, e^{-\mathrm{i}\Omega(t_{1}-t_{2})}
\sum\limits_{n}\frac{\psi_{n}(\textbf{r}_1)\overline{\psi}_{n}(\textbf{r}_2)}{\Omega-E_{n}(1-\mathrm{i}0)}
\qquad,
\end{eqnarray}
and the sum over $n$ in Eq. (\ref{7}) runs over the entire Dirac spectrum. The thermal photon propagator in coordinate space representation in the Feynman gauge \cite{SLP-QED, S-2020} is
\begin{eqnarray}
\label{8}
D_{\mu\nu}^{\beta}(x_{1}x_{2})=-\frac{g_{\mu\nu}}{\pi r_{12}}\int\limits_{-\infty}^{\infty}d\omega\, n_{\beta}(|\omega|)
\\\nonumber
\times
\mathrm{sin}(|\omega|r_{12})e^{-\mathrm{i}\,\omega(t_{1}-t_{2})}
\qquad,
\end{eqnarray}
where $n_{\beta}(\omega)=(\mathrm{exp}(\beta\omega)-1)^{-1}$
is the photon density number of BBR field, $ \beta=1/k_{\mathrm{B}}T $, and $r_{12}\equiv |\textbf{r}_1-\textbf{r}_2|$.

The first-order $S$-matrix element (\ref{6a}) contains the perturbation itself and corresponds to the Feynman diagram with the magnetic-field interaction only (the tiny line with the cross in Fig.~\ref{fig3}). The resulting energy shift is given by Eq. (\ref{g1}). The second-order $S$-matrix element (\ref{6b}) gives the one-loop self-energy correction, which was described in detail for the hydrogen atom in \cite{SLP-QED}, and its real part yields an expression for the thermal ac-Stark shift of energy levels. Finally, Eqs. (\ref{6c})--(\ref{6d}) match the three Feynman graphs describing the thermal self-energy radiative corrections to the magnetic-field interaction, which is the topic of our present research. 

After substitution of Eqs. (\ref{6a})--(\ref{6d}) into Eq. (\ref{5}), the total energy shift of the third order in $ e $ can be written as a sum of vertex (Fig. \ref{fig3} (a)) and wave-function (Fig. \ref{fig3} (b), (c)) contributions $ \Delta E_{a}=\Delta E_{a}^{\mathrm{ver}}+\Delta E_{a}^{\mathrm{wf}}$. According to \cite{lindgren}, the vertex contribution reads
\begin{eqnarray}
\label{9}
\Delta E_{a}^{\mathrm{ver}}=\mathrm{Re}\lim\limits_{\eta\rightarrow 0+}\frac{3\mathrm{i}\eta}{2}\langle \Phi_{a}^0 |\hat{S}_{\eta}^{(3)}| \Phi_{a}^0 \rangle
_{\mathrm{(a)}}
\,,
\end{eqnarray}
while the wave-function contribution is given by
\begin{eqnarray}
\label{10}
\Delta E_{a}^{\mathrm{wf}}=\mathrm{Re}\lim\limits_{\eta\rightarrow 0+}\frac{3\mathrm{i}\eta}{2}
\left(
\langle \Phi_{a}^0 |\hat{S}_{\eta}^{(3)}| \Phi_{a}^0 \rangle
_{\mathrm{(b)}} 
\right.
\\\nonumber
-
\left.
\langle \Phi_{a}^0 |\hat{S}_{\eta}^{(1)}| \Phi_{a}^0 \rangle
\langle \Phi_{a}^0 |\hat{S}_{\eta}^{(2)}| \Phi_{a}^0 \rangle
\right)
\,.
\end{eqnarray}

Performing integration over time variables and taking the limit $ \eta\rightarrow 0+ $ \cite{LSP-sep}, we find for the vertex part,
\begin{eqnarray}
\label{11}
\Delta E_{a}^{\mathrm{ver}}
=-\frac{e^3 }{\pi}\mathrm{Re}
\sum\limits_{\pm}
\sum\limits_{n,m}
\int\limits_{0}^{\infty}
d\omega n_{\beta}(\omega)\qquad
\\\nonumber
\times
\frac{\langle am| \frac{\alpha^{\mu}_{1}\alpha_{3\mu}}{r_{13}} \sin(\omega r_{13}) |na\rangle
\langle n| \bm{\alpha}\textbf{A}|m \rangle 
}{(E_{a}\pm\omega-E_{n}(1-\mathrm{i}0))(E_{a}\pm\omega-E_{m}(1-\mathrm{i}0))}
\,,
\end{eqnarray}
where $\sum\limits_{\pm}$ denotes the sum of two contributions with $+$ and $-$ in energy denominators. Similar evaluation for the wave-function part gives
\begin{eqnarray}
\label{12}
\Delta E_{a}^{\mathrm{wf}}
=-\frac{2e^3 }{\pi}\mathrm{Re}
\sum\limits_{\pm}\int\limits_{0}^{\infty}
d\omega n_{\beta}(\omega)\qquad
\\
\nonumber
\times
\left[
\sum\limits_{\substack{n,m\\m\neq a}}
\frac{\langle an| \frac{\alpha^{\mu}_{1}\alpha_{2\mu}}{r_{12}} \sin(\omega r_{12}) |nm\rangle
\langle m| \bm{\alpha}\textbf{A}|a \rangle 
}{(E_{a}\pm\omega-E_{n}(1-\mathrm{i}0))(E_{a}-E_{m})}
\right.
\\\nonumber
\left.
-\frac{1}{2}
\sum\limits_{\substack{n}}
\frac{\langle an| \frac{\alpha^{\mu}_{1}\alpha_{2\mu}}{r_{12}} \sin(\omega r_{12}) |na\rangle
\langle a| \bm{\alpha}\textbf{A}|a \rangle 
}{(E_{a}\pm\omega-E_{n}(1-\mathrm{i}0))^2}
\right]
\,,
\end{eqnarray}
where, the `reference-state' contribution, $ m=a $, is presented by the second term \cite{Blunden}.  
Unlike the ordinary `zero-vacuum' corrections, expressions (\ref{11}) and (\ref{12}) do not have ultraviolet divergences due to the natural cut-off provided by the distribution function $ n_{\beta}(\omega) $. 

In the nonrelativistic limit ($ \alpha Z\ll 1 $) the integrand $ \sin(\omega r)/r $ can be expanded into Taylor series, leading to
\begin{eqnarray}
\label{expansion}
\langle ab| \frac{\alpha^{\mu}_{i}\alpha_{j\mu}}{r_{ij}} \sin(\omega r_{ij}) |cd\rangle \approx
\omega\delta_{ac}\delta_{bd}
\\\nonumber
-\frac{\omega^3}{6}\left(\langle a|r_{i}^2|c\rangle \delta_{bd} 
+ \delta_{ac}\langle b|r_{j}^2|d\rangle\right)
\\\nonumber
-\omega \langle a|\textbf{p}|c\rangle  \langle b|\textbf{p}|d\rangle
+\frac{\omega^3 }{3}\langle a|\textbf{r}|c\rangle  \langle b|\textbf{r}|d\rangle
\,,
\end{eqnarray}
where the relation $ r_{ij}^2=r_{i}^2+r_{j}^2-2\textbf{r}_{i}\textbf{r}_{j}$ is taken into account. 

Then substituting Eq. (\ref{expansion}) into Eqs. (\ref{11}) and (\ref{12}) and using a transformation between the velocity and length forms of the electric dipole matrix elements (see Appendix A for details), we obtain the nonrelativistic limit of the vertex and wave-function contributions,
\begin{align}
\label{13}
\Delta E_{a}^{\mathrm{ver}}
=-\frac{2Be^2\mu_{B}}{3\pi}\mathrm{Re}
\sum\limits_{\pm}
\sum\limits_{n,m}
\int\limits_{0}^{\infty}
d\omega \omega^3 n_{\beta}(\omega)\qquad
\\\nonumber
\times
\frac{\langle a| \textbf{r} |n\rangle \langle n|v_{z}|m \rangle \langle m|\textbf{r}|a\rangle
}{(E_{a}\pm\omega-E_{n}(1-\mathrm{i}0))(E_{a}\pm\omega-E_{m}(1-\mathrm{i}0))}
\,,
\end{align}
\begin{align}
\label{14}
\Delta E_{a}^{\mathrm{wf}}
=-\frac{4Be^2\mu_{B}  }{3\pi}
\mathrm{Re}\sum\limits_{\pm}
\int\limits_{0}^{\infty}
d\omega n_{\beta}(\omega)
\omega^3\qquad
\\\nonumber
\times
\left[
\sum\limits_{\substack{n,m\\m\neq a}}
\frac{\langle a|\textbf{r}|n\rangle \langle n|\textbf{r}|m\rangle \langle m| v_{z}|a \rangle 
}{(E_{a}\pm\omega-E_{n}(1-\mathrm{i}0))(E_{a}-E_{m})}
\right.
\\\nonumber
-
\left.
\frac{1}{2}
\sum\limits_{n}
\frac{\langle a|\textbf{r}|n\rangle \langle n|\textbf{r}|a\rangle \langle a| v_{z}|a \rangle 
}{(E_{a}\pm\omega-E_{n}(1-\mathrm{i}0))^2}
\right]
\,,
\end{align}
where the notation $v_{z}\equiv j_{z}+s_{z}$ is introduced. Here, the wave functions in the matrix elements of the numerators should be understood as solutions of the Schr\"odinger equation for H-like ions with point-like nucleus and an arbitrary nuclear charge $ Z $. Then, the nonrelativistic energies are independent on orbital and total angular momenta, i.e. $ E_{a}\equiv E_{n_{a}} $, and the dependence of the energy shift on the angular momenta enters via the angular reduction of nonrelativistic operators in Eqs. (\ref{13}) and (\ref{14}). The scalar product of two coordinate operators in Eqs. (\ref{13}), (\ref{14}) can decomposed via spherical components as follows $\textbf{r}\textbf{r}'=\sum_{q=0,\pm 1}(-1)^{q}r_{q}r'_{-q}$. In the $ lsjm_{j} $ coupling scheme, it can be evaluated with
\begin{eqnarray}
\label{16}
\langle n'l's'j'm_{j'}|r_{q}|nlsjm_{j}\rangle = 
\delta_{s's}
(-1)^{j'+j+l'+s+1-m_{j'}}
\\\nonumber
\times
\begin{pmatrix}
j'     & 1 & j     \\
-m_{j'} & q & m_{j}
\end{pmatrix}
\Pi_{j'}\Pi_{j}\Pi_{l'}\Pi_{l}
\begin{Bmatrix}
l' & j' & s\\
j  & l  & 1
\end{Bmatrix}
\\\nonumber
\times
(-1)^{l'}
\begin{pmatrix}
l'     & 1 & l    \\
0     & 0 & 0
\end{pmatrix}
\int\limits_{0}^{\infty}dr r^3R_{n'l'}(r)R_{nl}(r)
\,,
\end{eqnarray}
where $ R_{nl}(r) $ is the radial part of the Schr\"odinger wave function, $ \Pi_{a}=\sqrt{(2a+1)} $  and the usual notation for $3j$-symbols are employed \cite{VMK}. For the operators $ j_{z} $ and $ s_{z} $ we have
\begin{eqnarray}
\label{17}
\langle n'l's'j'm_{j'}|j_{z}|nlsjm_{j}\rangle = 
\\\nonumber
\times
\delta_{n'n}\delta_{l'l}\delta_{s's}\delta_{j'j}
\sqrt{\frac{j(j+1)(2j+1)}{2j'+1}} C_{jm10}^{j'm_{j'}}
\,,
\end{eqnarray}
\begin{eqnarray}
\label{18}
\langle n'l's'j'm_{j'}|s_{z}|nlsjm_{j}\rangle = 
\delta_{n'n}\delta_{l'l}\delta_{s's}
\\\nonumber
\times
(-1)^{2j'+l+s+1-m_{j'}}
\Pi_{j'}\Pi_{j}\sqrt{s(s+1)(2s+1)}
\\\nonumber
\times
\begin{pmatrix}
j'          & 1 & j     \\
-m_{j'}     & 0 & m_{j}
\end{pmatrix}
\begin{Bmatrix}
s      & l & j     \\
j'     & 1 & s
\end{Bmatrix}
\,.
\end{eqnarray}
After substitution of Eqs. (\ref{17}) and (\ref{18}) into Eqs. (\ref{13}) and (\ref{14}) the terms with the off-diagonal matrix elements of $v_{z}$ do vanish. 


For hydrogen-like ions with nuclear charge $ Z $, the parametric estimate of Eqs. (\ref{13}) and (\ref{14}) can be found by taking into account that in relativistic units $ r\sim (m_{e}\alpha Z)^{-1} $, $ E_{a}\sim m_{e}(\alpha Z)^2 $ and
$\int\limits_{0}^{\infty}d\omega\omega^{k}n_{\beta}(\omega)\sim(k_{B}^{\mathrm{\;r.u.}}T)^{k+1}$. Then the $g$-factor correction (\ref{corr}) is parametrized as follows
\begin{eqnarray}
\label{est1}
\Delta g_{a}\sim \frac{(k_{B}T)^4_{\mathrm{r.u.}}}{\alpha^5 m_{e}^4Z^6}
\,.
\end{eqnarray}
In particular, the estimation Eq. (\ref{est1}) is valid for the $ a=1s $ state, when the summation runs over the $ np $ states and the energy difference in the denominators of Eqs. (\ref{13}) and (\ref{14}) is always of the order $ m_{e}(\alpha Z)^2 $. However, for the $ n_{a}l_{a} $ states with $ n_{a} \geq 2 $ (e.g. $a=2s$), the dominant contribution in the sum over $ n $ arises from the terms with $ E_{n}=E_{n_{a}} $. In this case, the following parametrization is valid
\begin{eqnarray}
\label{est2}
\Delta g_{a}\sim \frac{(k_{B}T)^2_{\mathrm{r.u.}}}{\alpha m_{e}^2Z^{2}}
\,.
\end{eqnarray}
It should be noted that the quadratic behavior in temperature given by the formula (\ref{est2}) was predicted earlier for thermal radiative corrections to the free electron $g$-factor in \cite{DHR, Jun1991, Elmfors}.

Estimation (\ref{est2}) suggests that for $2s$ and $2p$ states, which are already accessible for high-precision $g$-factor measurements, the thermal self-energy correction can be relatively large and thus should be carefully investigated.
Below we present the numerical calculations of this correction for $1s$, $2s$, $2p_{1/2}$, and $2p_{3/2}$ states according to the formulas obtained above.



\section{Results and discussion}
\label{s2}

The one-loop thermal self-energy correction to the bound-electron $g$-factor is given by the formulas (\ref{13}) and (\ref{14}). To sum up the entire spectrum in Eq. (\ref{13}), we use the Schr\"odinger finite basis set constructed within the nonrelativistic B-spline approach, following the ideas presented in Refs.~\cite{sapirstein:1996:5213, DKB}. In our calculations, we also modify the nonrelativistic energies of $2s$ and $2p$ states by including the Lamb shift correction \cite{lambZ}. Within the framework of the Line Profile approach, this can be performed by the accounting of an infinite series of self-energy and vacuum-polarization insertions into the internal electron lines of diagrams in Fig.~(\ref{fig3}). Arising geometric progression leads to the appearance of the corresponding Lamb shifts and levels widths in the energy denominators \cite{Andr}. As was shown in \cite{SLP-QED, Reexamine}, inclusion of the Lamb shift slightly changes the BBR-induced ac-Stark shift in hydrogen. Fine splitting is also taken into account for the $ 2p_{1/2} $ and $ 2p_{3/2} $ states, where it is important. 

Numerical results for the one-loop thermal correction to the bound-electron $g$-factor according to Eqs. (\ref{13}) and (\ref{14}) are given in Table I for H-like ions with $ Z=1,\,2,\,6 $.
\begin{table}
\caption{Numerical values of thermal one-loop self-energy corrections to the $g$-factor of the bound electron.}
\begin{tabular*}{\columnwidth}{ c c r r r }
\hline
&  T, Kelvin 
& \multicolumn{1}{c}{77} 
& \multicolumn{1}{c}{300}
& \multicolumn{1}{c}{1000}
\\
\hline
\hline
 &  &  & \multicolumn{1}{c}{$ Z=1 $} &  
 \\
&  $\Delta g[{1s_{1/2}}]$ &   $3.42\times 10^{-20}$  &  $8.34\times 10^{-18}$  &   $1.03\times 10^{-15}$ 
\\
&  $\Delta g[{2s_{1/2}}]$ &   $2.58\times 10^{-13}$  &  $3.92\times 10^{-12}$  &  $4.37\times 10^{-11}$ 
\\
&  $\Delta g[{2p_{1/2}}]$ &  $1.94\times 10^{-13}$   &  $2.94\times 10^{-12}$  &  $3.24\times 10^{-11}$    
\\
&  $\Delta g[{2p_{3/2}}]$ &   $9.68\times 10^{-14}$  &  $1.47\times 10^{-12}$  &  $1.62\times 10^{-11} $ 
\\
 &  &  & \multicolumn{1}{c}{$ Z=2 $} &  
 \\ 
&  $\Delta g[{1s_{1/2}}]$ &   $5.34\times 10^{-22}$  &  $1.30\times 10^{-19}$  &   $1.61\times 10^{-17}$ 
\\
&  $\Delta g[{2s_{1/2}}]$ &  $6.45\times 10^{-14}$   &  $9.79\times 10^{-13}$  &   $1.09\times 10^{-11}$ 
\\
&  $\Delta g[{2p_{1/2}}]$ &   $ 4.84\times 10^{-14} $   & $7.35\times 10^{-13}$    &  $8.16\times 10^{-12}$ 
\\
&  $\Delta g[{2p_{3/2}}]$ &  $2.35\times 10^{-14}$  &  $3.66\times 10^{-13}$   &  $4.78\times 10^{-12}$ 
\\
 &  &  & \multicolumn{1}{c}{$ Z=6 $} &  
 \\
&  $\Delta g[{1s_{1/2}}]$ &  $ 7.33\times 10^{-25}$ & $1.79\times 10^{-22}$ &  $2.21\times 10^{-20} $   
\\
&  $\Delta g[{2s_{1/2}}]$ &  $7.18\times 10^{-15}$   &  $1.09\times 10^{-13}$   &  $1.21\times 10^{-12}$ 
\\
&  $\Delta g[{2p_{1/2}}]$ &  $5.38\times 10^{-15}$    &   $8.16\times 10^{-14}$   &  $9.07\times 10^{-13}$ 
\\
&  $\Delta g[{2p_{3/2}}]$ &  $2.52\times 10^{-16}$  &  $-2.32\times 10^{-14}$   &  $1.97\times 10^{-13}$ 
\\
\hline
\end{tabular*}
\end{table}
For the ground $1s$ state of all the considered ions, we conclude that the effect is insignificant at cryogenic and room temperatures at the current level of experimental accuracy. However, it is important to consider within the framework of such demanding scenarios as search for new physics \cite{5th}, in assumption that the required experimental progress is possible. 

The situation is different for the $g$-factor of $n=2$ states states. In particular, for the $2s_{1/2}$ and $2p_{1/2}$ levels the one-loop thermal corrections at $T=300$ K are on the level of precision anticipated in the forthcoming experiments \cite{blaum:21:qst}. This contribution is comparable to various previously investigated higher-order effects, such as the bound-state two-loop QED contributions \cite{twoloopG}, in particular, the two-loop virtual light-by-light scattering \cite{czarnecki:2018:043203, czarnecki:2020:050801, PhysRevA.103.L030802}. We note that the results for the $2s$ and $2p$ states are relevant not only for the excited states of H-like ions, but also for the ground states of Li- and B-like ions, respectively. In few-electron ions the values of the correction may differ significantly due to the screening effects, but the order-of-magnitude estimation is still valid. Meanwhile, the Li- and B-like ions are important for various proposed investigations, in particular, for determination of the fine structure constant \cite{shabaev:2006:253002,volotka:2014:023002,yerokhin:2016:100801}.

An equally interesting aspect of such effects is the search for variations of the fundamental constants: the fine structure constant, the gravitational constant, and the proton-to-electron mass ratio \cite{DIRAC1937,flambaum:2009:597,safronova:19:ap}. Various phenomenological limits or evidence reported from astrophysics, cosmology and laboratory experiments are still under discussion. This interest is periodically spurred by the `positive' results concerning the variation of the fine-structure constant $\alpha$, see, for example, \cite{Webb}. Nevertheless, the reported results should be interpreted with care: the laboratory experiments has demonstrated especially fast progress, requiring the consideration of ever smaller effects, see \cite{Labzowsky_2007} and discussion therein. Assuming an accurate comparison between laboratory and astrophysical experiments, the correction values are listed in Table~I for higher temperatures.

Special attention should be paid to the case of $ Z=6 $ H-like ion. The experimental measurement of $g$-factor in $^{12}\mathrm{C}^{\mathrm{5+}}$ ions allows the determination of the electron mass value with a record accuracy \cite{nature2014}. The achieved level of experimental accuracy leads to the electron mass determination with a relative error of about $ 10^{-11} $. The corrections obtained in present work are close to this value and, can be considered as possible source for future experimental improvements.


\section{Conclusion}

The bound-electron $g$-factor is a subject of high-precision measurements and theoretical calculations, which jointly are able to bring various results of fundamental importance. The effect of black-body radiation was considered previously for binding energies showing its importance in Rydberg atoms, atomic clocks, and frequency standards. In this work, we have investigated one-loop thermal correction for $1s$, $2s$, $2p_{1/2}$, and $2p_{3/2}$ states of light H-like ions. Currently available experimental values are found to be insensitive to blackbody radiation. However, the potential importance of this effect for $n=2$ states in future investigations is demonstrated, including determination of the fundamental constants and search for new physics.


\section{Acknowledgements}
This work was supported by Russian Foundation for Basic Research (grants No. 20-02-00111 and 19-02-00974). 

\appendix
\renewcommand{\theequation}{A\arabic{equation}}
\setcounter{equation}{0}

\section*{Appendix A: Relation between length and velocity forms of matrix elements}
In this Appendix we present the relations between length and velocity matrix elements employed in section \ref{s1} for derivation of the nonrelativistic limit from general relativistic equations. The Taylor expansion Eq. (\ref{expansion}) leads to the product of electron momenta matrix elements in the first term which can be converted to the product of space vector matrix elements. Then, in conjunction with the second term of Eq. (\ref{expansion}), we can obtain the so-called length-form of expansion. To do this, we first recall the well-known commutation relation $ \textbf{p}=\mathrm{i}[\hat{H},\textbf{r}] $, where $ \hat{H} $ is the operator of the total Hamiltonian of the system. In the absence of external field for the nonresonant transition from the bound state with the energy $ E_{a} $ to the virtual intermediate state with the energy $ E_{n} $ the above commutation relation takes the extended form:
\begin{eqnarray}
\label{ext}
\textbf{p}=\mathrm{i}[\hat{H}-E_{a}\pm\omega,\textbf{r}] 
,
\end{eqnarray} 
where sign $\pm $ depends on absorption or emission process under consideration, respectively. Then, taking into account Eq. (\ref{ext}), the following expression can be easily proven
\begin{widetext}
\begin{align}
\label{exp1}
-\omega
\sum\limits_{\pm}&\sum\limits_{n,m\neq a}\frac{\langle a|\textbf{p}| n\rangle\langle n |\textbf{p}|m\rangle \langle m|\hat{T}_{0} |a\rangle}{(E_{n}-E_{a}\pm\omega)(E_{m}-E_{a})}
\\\nonumber
&=
-\omega
\sum\limits_{\pm}\sum\limits_{m\neq a}
\left[
(E_{m}-E_{a}\pm\omega)(\pm\omega)\sum\limits_{n}\frac{\langle a|\textbf{r}| n\rangle\langle n |\textbf{r}|m\rangle}{E_{n}-E_{a}\pm\omega}
-(E_{m}-E_{a})\langle a|r^2|m\rangle + 3\delta_{am}
\right]\frac{\langle m|\hat{T}_{0}|a\rangle}{E_{m}-E_{a}}
\\\nonumber
&=
-\omega
\sum\limits_{\pm}\sum\limits_{n,m\neq a}
\left[
(\pm\omega)\frac{\langle a|\textbf{r}| n\rangle\langle n |\textbf{r}|m\rangle \langle m|\hat{T}_{0} |a\rangle}{E_{n}-E_{a}\pm\omega}
+
\omega^2\frac{\langle a|\textbf{r}| n\rangle\langle n |\textbf{r}|m\rangle \langle m|\hat{T}_{0} |a\rangle}{(E_{n}-E_{a}\pm\omega)(E_{m}-E_{a})}
\right]
+\omega\sum\limits_{m}\langle a|r^2|m \rangle\langle m|\hat{T}_{0} |a\rangle
\\\nonumber
&=
-\omega
\sum\limits_{\pm}
\left[
(\pm\omega)\sum\limits_{n}\frac{\langle a|\textbf{r}| n\rangle\langle n |\textbf{r}\hat{T}_{0} |a\rangle}{E_{n}-E_{a}\pm\omega}
+
\omega^2\frac{\langle a|\textbf{r}| n\rangle\langle n |\textbf{r}|m\rangle \langle m|\hat{T}_{0} |a\rangle}{(E_{n}-E_{a}\pm\omega)(E_{m}-E_{a})}
\right]
+\omega\langle a|r^2\hat{T}_{0} |a\rangle
,
\end{align}
\end{widetext}
where $\hat{T}_{0}$ is an arbitrary one-electron Hermitian scalar operator.
The relation given by Eq. (\ref{exp1}) is used in the reducible part of the energy shift Eq. (\ref{10}), when the first terms in the expansion Eq. (\ref{expansion}) are converted into the length-form. First two terms in Eq. (\ref{exp1}) can be neglected since they are exactly cancelled by the similar contributions arising from the vertex contribution (\ref{11}). 

For the latter one can write the following expression:
\begin{widetext}
\begin{align}
\label{exp2}
-\omega\sum\limits_{\pm}&\sum\limits_{nm}\frac{\langle a|\textbf{p}| n\rangle \langle n|\hat{T}_{0} |m\rangle \langle m |\textbf{p}|a\rangle }{(E_{n}-E_{a}\pm\omega)(E_{m}-E_{a}\pm\omega)}
=
-\omega
\sum\limits_{\pm}\sum\limits_{nm}
\left[
(\mp\omega)
\frac{
\langle a|\textbf{r}|n\rangle \langle n|\hat{T}_{0} |m\rangle \langle m|\textbf{r}|a\rangle}{E_{n}-E_{a}\pm \omega}
\right.
\\\nonumber
&\left.
+
(\mp\omega)\frac{\langle a|\textbf{r}|n\rangle \langle n|\hat{T}_{0} |m\rangle\langle m|\textbf{r}|a\rangle}{E_{m}-E_{a}\pm \omega}
+2\langle a|\textbf{r}|n\rangle \langle n|\hat{T}_{0} |m\rangle \langle m|\textbf{r}|a\rangle
+\omega^2\frac{
\langle a|\textbf{r}|n\rangle \langle n|\hat{T}_{0} |m\rangle \langle m|\textbf{r}|a\rangle}{(E_{n}-E_{a}\pm \omega)(E_{m}-E_{a}\pm \omega)}
\right]
\\\nonumber
&=
-\omega
\sum\limits_{\pm}
\left[
2(\mp\omega)
\sum\limits_{n}
\frac{
\langle a|\textbf{r}|n\rangle \langle n|\textbf{r}\hat{T}_{0}|a\rangle}{E_{n}-E_{a}\pm \omega}
+\omega^2
\sum\limits_{nm}
\frac{
\langle a|\textbf{r}|n\rangle \langle n|\hat{T}_{0} |m\rangle \langle m|\textbf{r}|a\rangle}{(E_{n}-E_{a}\pm \omega)(E_{m}-E_{a}\pm \omega)}
\right]
-2\omega\langle a|r^2\hat{T}_{0}|a\rangle
,
\end{align}
\end{widetext}
which is also obtained with the use of Eq. (\ref{ext}) and the sum rule $ \sum\limits_{n} |n\rangle \langle n| =1 $. After the substitution of Eqs. (\ref{exp1})-(\ref{exp2}) into Eqs. (\ref{11})-(\ref{expansion}) the terms linear in $ \omega $ in the total energy shift vanish, since the wave-function contribution is doubled relative to the vertex one. Then it is easy to see that only terms proportional to $ \omega^3 $ survive and we arrive at final nonrelativistic equations for energy shift given by Eqs. (\ref{13}), (\ref{14}).

\bibliography{mybibfile}

\end{document}